\documentclass[twocolumn]{aastex62}

\usepackage{apjfonts}

\shorttitle{Planetary Nebula JaFu 1 in Palomar 6}
\shortauthors{Bond et al.}

\newcommand{\Ha}{H$\alpha$}

\newcommand{\kms}{{\>\rm km\>s^{-1}}}

\def\oiii{\ion{O}{3}}

\def\Gaia{{\it Gaia}}

\newcommand{\HST}{{\it HST}}

\begin{document}


\title{Testing Cluster Membership of Planetary Nebulae with High-Precision Proper Motions. I\null. \\ \HST\/ Observations of JaFu~1 Near the Globular Cluster Palomar 6}

\author[0000-0003-1377-7145]{Howard E. Bond}
\affil{Department of Astronomy \& Astrophysics, Pennsylvania State University, University Park, PA 16802, USA}
\affil{Space Telescope Science Institute, 
3700 San Martin Dr.,
Baltimore, MD 21218, USA}

\author[0000-0003-3858-637X]{Andrea Bellini}
\affil{Space Telescope Science Institute, 
3700 San Martin Dr.,
Baltimore, MD 21218, USA}

\author[0000-0001-6008-1955]{Kailash C. Sahu}
\affil{Space Telescope Science Institute, 
3700 San Martin Dr.,
Baltimore, MD 21218, USA}
\affil{Eureka Scientific Inc., 2542 Delmar Avenue, Suite 100, Oakland, CA 94602, USA}

\correspondingauthor{Howard E. Bond}
\email{heb11@psu.edu}

\begin{abstract}

If a planetary nebula (PN) is shown to be a member of a star cluster, we obtain important new constraints on the mass and chemical composition of the PN's progenitor star, which cannot be determined for PNe in the field. Cluster membership can be tested by requiring the projected separation between the PN and cluster to be within the tidal radius of the cluster, and the objects to have nearly identical radial velocities (RVs) and interstellar extinctions, and nearly identical proper motions (PMs). In an earlier study, we used PMs to confirm that three PNe, which had already passed the other tests, are highly likely to be members of Galactic globular clusters (GCs). For a fourth object, the PN JaFu\,1, which lies in the Galactic bulge near the GC Palomar~6 on the sky and has a similar RV, the available PM measurement gave equivocal results. We have now obtained new high-resolution images of the central star of JaFu\,1 with the {\it Hubble Space Telescope\/} (\HST) which, combined with archival \HST\/ frames taken 14 and 16 years earlier, provide a high-precision PM\null. Unfortunately, we find that the PM of the central star differs from that of the cluster with high statistical significance, and thus is unlikely to be a member of Palomar~6. Nevertheless, JaFu\,1 is of astrophysical interest because its nucleus appears to be a member of the rare class of ``EGB\,6-type'' central stars, which are associated with compact emission-line knots.

\null\vskip 0.2in

\end{abstract}



\section{Introduction: Planetary Nebulae in Open and Globular Star Clusters \label{sec:intro} }

Planetary nebulae (PNe) mark the final, rapid transition of a low- or
intermediate-mass star from the asymptotic giant branch (AGB) to the top of the
white-dwarf cooling sequence. PNe provide information on such topics as late stages of stellar evolution,
stellar nucleosynthesis, mixing processes that bring processed material to the
surface, and the ejection of these elements into the
interstellar medium. { For overviews of PNe see, for example, \citet{Kwitter2014}, \citet{Weidmann2020}, and \citet{Kwitter2022}. }


Studies of PNe were long hampered by uncertainties in the distances to Galactic PNe.
Thus basic information such as the luminosities of the central stars and the
physical sizes and masses of the nebulae was poorly constrained. This situation has been improved considerably in the last few years by the availability of trigonometric parallaxes for many planetary-nebula nuclei (PNNi) from the \Gaia\/ mission \citep[e.g.,][and references therein]{Bucciarelli2023, Chornay2023, Hernandez2024}. 

However, even if a precise
distance is available for a PN, the mass and composition of its central star's progenitor remain
unknown. This means that finding and verifying PNe that are {\it members of star clusters}
is crucially important. For these objects, the progenitor star's mass and
metallicity can be determined from the cluster's main-sequence turnoff
and metal content; the absolute luminosities of the PN and the PNN can be found using the cluster's
distance via main-sequence fitting; and the mass of the PNN can be deduced from
theoretical core-mass\slash luminosity relations---thus providing a point on the important initial-final mass relation \citep[e.g.,][]{Cummings2018, Barnett2021, Marigo2022}. Additionally, the chemical
composition of the PN, along with the progenitor mass, provide information about the dredge-up of processed
material from the stellar interior---theoretically predicted to depend strongly on the
initial mass of the progenitor star \citep[e.g.,][]{Karakas2016}.

Unfortunately, only a very few PNe have been shown to be probable members of star clusters. Among open clusters (OCs) a convincing case is IPHASX J055226.2+323724
(PN\,G177.6+03.0). This PN has recently been confirmed as a member of the OC M37 (NGC~2099), based on the \Gaia\/ proper motion (PM) of its central star, its nebular radial velocity (RV), and its interstellar extinction, all of which agree with those of the cluster \citep{Griggio2022, FragkouM37_2022, WernerM372023}. There are over a dozen other Galactic PNe that
lie near OCs on the sky, but nearly all fail the membership tests of
having the same PM, RV, and extinction as the cluster (see detailed discussion and references in \citealt{Frew2016} and \citealt{Davis2019}). Only two Galactic PNe remain at present as strong candidate members of OCs to add to the case of the M37 PN\null .
These are  PHR~J1315$-$6555 in the OC ESO 96-SC04 = AL~1 \citep{Parker2011, FragkouPHR2019}, and BMP~J1613$-$5406 in NGC~6067 \citep{Frew2016, FragkouBMP2019, FragkouNGC6067_2022}.  In the Local Group, a PN that likely belongs to an OC in M31 was identified by 
\citet{Bond2015} and analyzed spectroscopically by \citet{Davis2019}.



In the case of globular clusters (GCs), there are three and possibly four known
cases of PNe that appear to belong to Galactic GCs \citep[][and references therein]{Jacoby1997, Bond2020}. Their central stars are likely to be
descended from interacting or merged binaries \citep[e.g.,][]{Bond2015, Jacoby2017, Sun2019}, since single low-mass post-AGB stars in these ancient clusters probably evolve too slowly to ionize their ejecta before they disperse. Our team \citep{Bond2020} obtained PMs of the central stars of these four PNe from \Gaia\/ Data Release~2 (DR2), or by measuring archival multi-epoch {\it Hubble Space Telescope\/} (\HST\/) frames. For three of them---the PNe in M15, M22, and NGC~6441---the PMs (as well as their RVs) provide strong support for cluster membership. For the fourth case, JaFu~1 near the GC Palomar~6 (hereafter Pal~6), whose central star is too faint for \Gaia, our PM measurement was less precise due to the relatively low quality and short time baseline of the available \HST\/ frames; nevertheless, our findings raised doubts about its cluster membership.

In order to investigate with more precision whether JaFu~1 is a member of Pal~6, we have obtained new \HST\/ images in order to significantly reduce the uncertainties in the PM of its central star, and we report the results here. In the same \HST\/ program, we have also imaged the nuclei of PHR~J1315$-$6555 and BMP~J1613$-$5406, and we will present our findings for these objects in future papers in this series (although the PM measurement for the latter target will have to await second-epoch \HST\/ observations to be obtained in 2026).

\section{J\MakeLowercase{a}F\MakeLowercase{u}~1: Candidate Member of Palomar~6 \label{sec:JaFu1} }

In the 1990s, G. Jacoby and L.~Fullton (see \citealt{Jacoby1997}) carried out ground-based narrow-band [\oiii] 5007~\AA\ imaging of 133 Galactic GCs, with the aim of finding PNe in the target clusters. Their survey resulted in discoveries of two new PNe that were candidate cluster members, designated JaFu~1 (PN\,G002.1+01.7; in Pal~6), and JaFu~2 (PN\,G353.5$-$05.0; in NGC~6441). As noted above, JaFu~2 appears to be a genuine member of its host cluster.

\begin{figure}
\centering
\includegraphics[width=\linewidth]{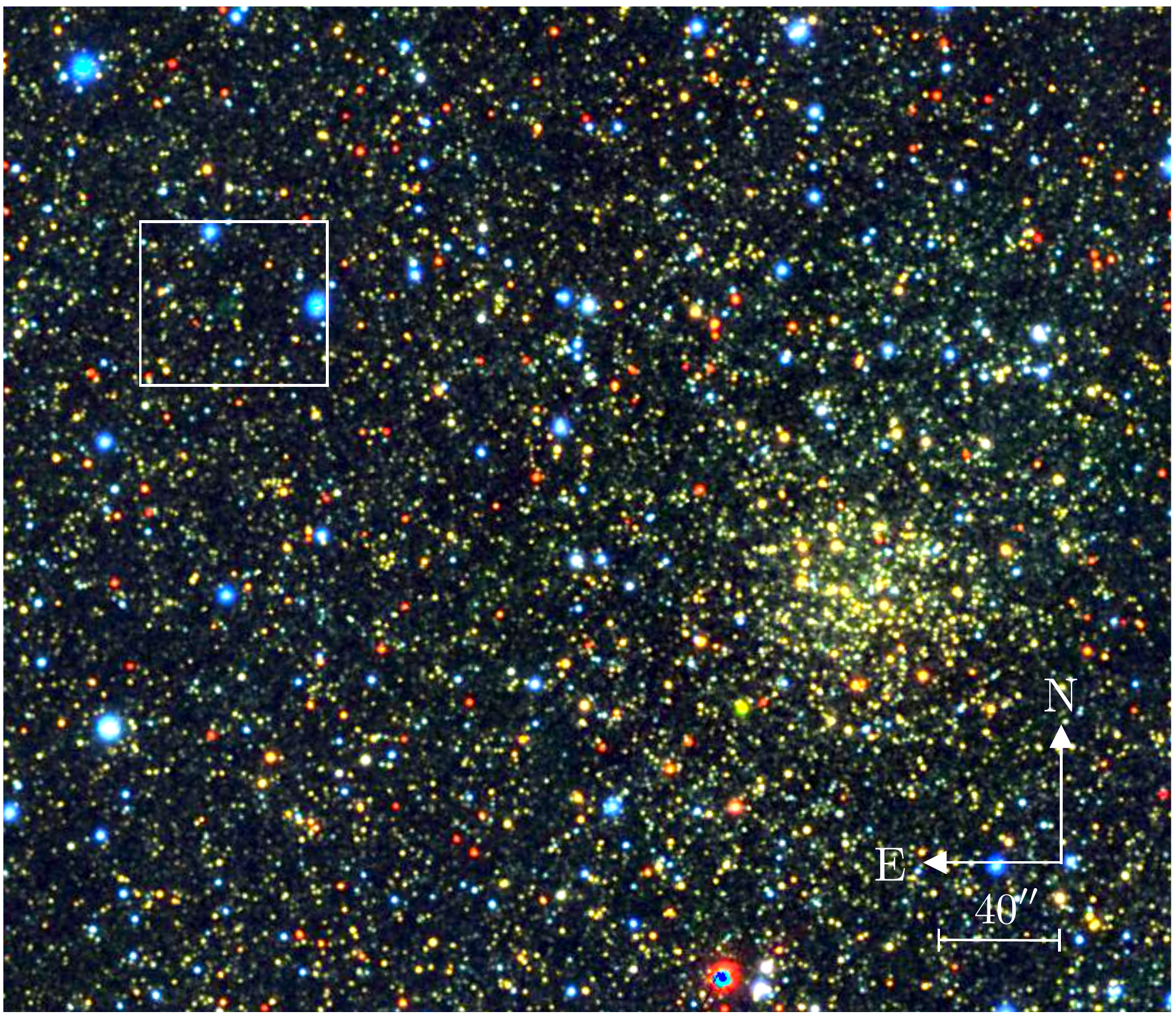}
\caption{
Color image created from PanSTARRS-1 $g$, $r$, and $i$ frames, showing the globular cluster Pal~6 (below right) and the location of the planetary nebula JaFu~1 (faintly visible at the center of the white rectangle at upper
left). North is at the top, east on the left. Height of frame is $320''$. 
\label{fig:panstarrs} }
\end{figure}

Pal~6 is a faint and heavily reddened GC, discovered by \citet{Abell1955} in the course of his examination of photographs from the Palomar Observatory Sky Survey (POSS). Lying in the Milky Way bulge at Galactic coordinates $l=2\fdg09, b=+1\fdg78$, Pal~6 is superposed on a rich star field. Figure~\ref{fig:panstarrs} shows a ground-based image of the cluster and its surroundings, created using the PanSTARRS-1 Image Access tool.\footnote{\url{https://ps1images.stsci.edu/cgi-bin/ps1cutouts}} Here the cluster lies at the lower right of the frame, and JaFu~1 is located at the center of the white rectangle at the upper left. 

JaFu~1 lies about $230''$ from the center of Pal~6, at a position where the stellar field is heavily dominated by bulge stars, as shown in Figure~\ref{fig:panstarrs}. However, this separation is still well within the cluster's tidal radius of $\sim$$500''$, calculated from data tabulated by \citet{Harris2010}.\footnote{The Harris compilation of GC properties, 2010 December version, is available online at \url{http://physwww.mcmaster.ca/~harris/mwgc.dat}} 


The distance of Pal~6 is about 7.2~kpc, according to an analysis of near-IR stellar photometry by \citet{Lee2002}. This has been updated more recently to $7.05^{+0.46}_{-0.43}$~kpc in a comprehensive compilation of GC distances by \citet{Baumgardt2021}. An independent estimate of the distance to JaFu~1 could in principle be obtained using the statistical relation between H$\alpha$ surface brightness and linear radius for PNe derived by \citet{Frew2016}; however, these authors actually used JaFu~1 as one of their calibrators for the method, assuming it to be a cluster member and adopting the \citet{Lee2002} cluster distance. In any case, since the position of JaFu~1 in the surface-brightness versus radius relations plotted by \citet{Frew2016} agrees well with other calibrators, the distances of the PN and the cluster are statistically similar. 


The question of JaFu~1's cluster membership has been reviewed in detail by \citet{Jacoby1997, Jacoby2017} and \citet{Bond2020}. We note here that the strongest evidence for membership comes from the emission-line RV measurement for JaFu~1 by \citet{Jacoby1997} of $+176\pm15\,\kms$. This agrees remarkably well with the RV of Pal~6 of $+176.3\pm1.5\,\kms$, determined from the mean of about a dozen individual member stars \citep{Baumgardt2019}. However, as \citet{Jacoby1997} noted, there is an appreciable velocity dispersion among Galactic-bulge stars. Thus there is a non-zero chance of a field PN having an RV similar to that of the cluster, estimated by \citet{Jacoby1997} to be about 3\%. An argument against membership is the large angular separation of PN and cluster; \citet{Jacoby1997} used the surface density of field PNe in the Galactic bulge to estimate a $\sim$15\% probability of a bulge PN lying this close to the cluster by chance. Potential negative evidence is that the interstellar reddening of the PN appears to be larger than that of Pal~6. \citet{Jacoby1997, Jacoby2017} derived reddening estimates for the PN of $E(B-V)=1.93$ and 1.67 from considerations of the nebular spectrum and the intrinsic color of the central star. The reddening of the cluster is lower, at $E(B-V)=1.46$ \citep{Harris2010}. However, inspection of POSS and PanSTARRS images of the field shows the extinction to be patchy and likely higher in the vicinity of the PN than it is for the cluster, so this membership test is not decisive. Lastly, as noted in the Introduction, \citet{Bond2020} measured a low-precision PM for the central star from \HST\/ images with a short time baseline; it was discordant with the mean PM of cluster member stars at about a 2$\sigma$ level.



\section{New and Archival \HST\/ Observations of J\MakeLowercase{a}F\MakeLowercase{u}~1}

Table~\ref{tab:exposures} lists the available \HST\/ images of JaFu~1. They were obtained during two visits in 2008 and 2010, and during our new visit in 2024.\footnote{Frames were downloaded from the Mikulski Archive for Space Telescopes at \url{https://archive.stsci.edu/}. All of the \HST\/ frames analyzed in this paper are available at \url{http://dx.doi.org/10.17909/d691-h991}.} The 2008 observations were made with the Wide Field Planetary Camera~2 (WFPC2), using its broad-band F555W ($V$) and F814W ($I$) filters, and its narrow-band F656N (H$\alpha$) filter. The WFPC2 frames were aimed at direct imaging of the faint PN and photometry of the central star, and are not optimal for precision stellar astrometry since they were obtained in the camera's low-resolution WF3 channel, and the telescope pointings were not dithered. The 2010 frames were obtained with the Wide Field Channel (WFC) of the Advanced Camera for Surveys (ACS), this time for imaging of the PN in the narrow-band F502N ([\oiii] 5007~\AA) filter, and again without dithering. 

\begin{deluxetable*}{lllccl}
\tablecaption{Log of {\it Hubble Space Telescope\/} Observations of JaFu~1 \label{tab:exposures} }
\tablehead{
\colhead{Date}
&\colhead{Camera and}
&\colhead{Filter}
&\colhead{Pixel Scale}
&\colhead{Exposures}
&\colhead{Program ID}\\
\colhead{}
&\colhead{Channel}
&\colhead{}
&\colhead{[mas pixel$^{-1}$]}
&\colhead{[s]}
&\colhead{and PI}
}
\startdata
2008 Mar 14 & WFPC2/WF3 & F555W & 100 & $2\times160$  & GO-11308 (O. De~Marco)\\
            &           & F656N & & $2\times500$  &  \\
            &           & F814W & & $2\times160$  &  \\
2010 Mar 14 & ACS/WFC   & F502N & 50 & $3\times796$  & GO-11558 (O. De~Marco)\\ 
2024 Feb 14 & WFC3/UVIS & F555W & 40 & $6\times120$  & GO-17531 (H.E.B.)\\
            &           & F656N & & $2\times300$  &  \\
\enddata
\end{deluxetable*}

For our new \HST\/ imaging in 2024, we used the UVIS (ultraviolet and visible light) channel of the Wide Field Camera~3 (WFC3) in F555W with a six-point dither pattern, along with two dithered exposures in F656N\null. To maximize observing efficiency, we employed a $2047\times2050$ pixel ($81''\times81''$) subarray.

\begin{figure}
\centering
\includegraphics[width=\linewidth]{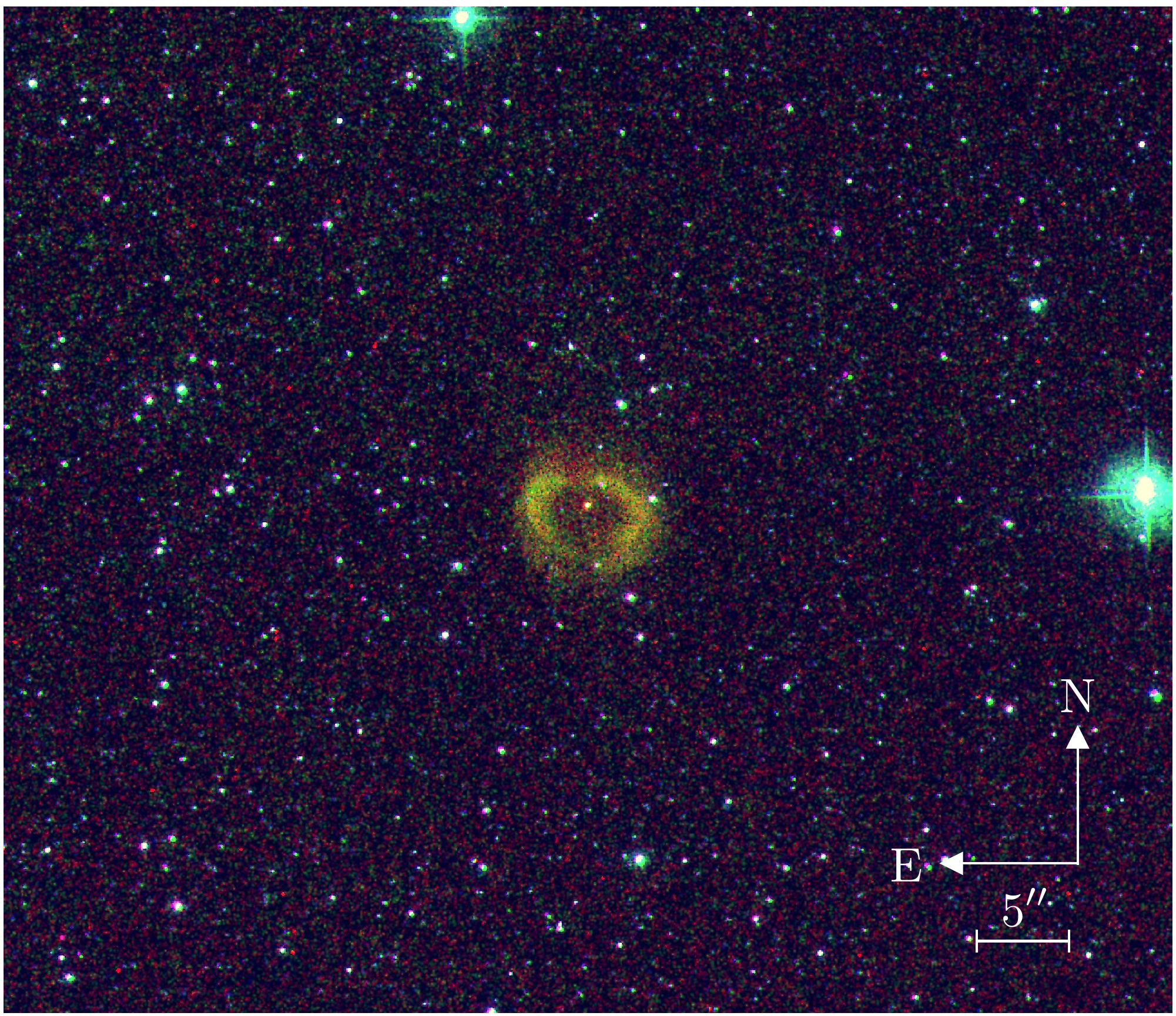}
\caption{
False-color \HST\/ image of the field centered on the planetary nebula JaFu~1, outlined by the white rectangle in Figure~\ref{fig:panstarrs}. Created from WFC3 frames in \Ha\ (red) and $V$ (blue), and an ACS frame in [\oiii] 5007~\AA\ (green). North is at the top, east on the left. Height of frame is $50''$.
\label{fig:HST_JaFu1} }
\end{figure}

Figure~\ref{fig:HST_JaFu1} shows a false-color image of JaFu~1 at \HST\/ resolution, prepared by combining the archival ACS frame in [\oiii] (green) with our WFC3 frames in \Ha\ (red) and $V$ (blue). Here we used the drizzle-combined images corrected for geometric distortion and charge-transfer efficiency (CTE), obtained from the Mikulski Archive pipeline (filename suffix {\tt\_drc}). The field shown in Figure~\ref{fig:HST_JaFu1} corresponds to the region enclosed in the white rectangle in Figure~\ref{fig:panstarrs}. Note that the [\oiii] image was obtained 14 years prior to the \Ha\ and $V$ data. We aligned the images such that JaFu~1 and its central star are at the same pixel location; as a consequence, a slight but noticeable misalignment of some of the field stars due to their intrinsic PMs can be seen.\footnote{Our figure can be compared with a color image of JaFu~1 prepared by \citet[][their Figure 1]{Jacoby2017} from the WFPC2 and ACS narrow-band images.}

The image of JaFu~1 shows an elliptical ring with diffuse edges, with minor and major axes of about $5\farcs5\times7\farcs3$ (corresponding to physical dimensions of $0.19\times0.25$~pc, at a distance of 7~kpc). The nucleus is conspicuously off-center to the north inside the surrounding ring.  


\section{Data Analysis \label{sec:dataanalysis} }


We computed absolute PMs of stars in the \HST\/ field following the
prescriptions given in \citet{Bellini14, Bellini18a}, and described in detail by \citet{Bond2020}. In brief, we
measured positions and magnitudes of neighbor-subtracted sources on the
pipeline-calibrated, unresampled exposures from the three epochs listed in Table~\ref{tab:exposures} (filename suffixes \texttt{\_flc}
for ACS and WFC3 with CTE corrections, \texttt{\_c0f} for WFPC2, which are not CTE-corrected).  We used a combination of
first- and second-pass reduction techniques based on the software
tools \texttt{hst1pass}\footnote{{\tt hst1pass} is available at \url{https://www.stsci.edu/hst/instrumentation/wfc3/software-tools}.} \citep{Anderson22a, Anderson22b} and
\texttt{KS2} (see \citealt{Bellini17}). These routines take
advantage of high-precision effective point-spread function (PSF) models
(\citealt{Anderson00, Anderson18, Bellini18b}) and
geometric-distortion solutions (\citealt{Anderson00, Anderson06,
Bellini09, Bellini11}). We defined a common reference-frame system
based on positions for the brighter stars in the field contained in the \textit{Gaia} Data Release~3\footnote{\url{https://vizier.cds.unistra.fr/viz-bin/VizieR-3?-source=I/355/gaiadr3}} (DR3; \citealt{Gaia2016, Gaia2023}) catalog. We then adjusted all
single-exposure source positions onto this frame by means of general
six-parameter linear transformations, using a local network of neighboring
sources around each star. Finally, for each source we determined its PM by fitting the slope of its reference-frame-transformed positions as a function of exposure epoch, via an
iterative central-overlap method, as described in \citet{Bellini14}.

\begin{figure*}
\centering
\includegraphics[width=0.8\linewidth]{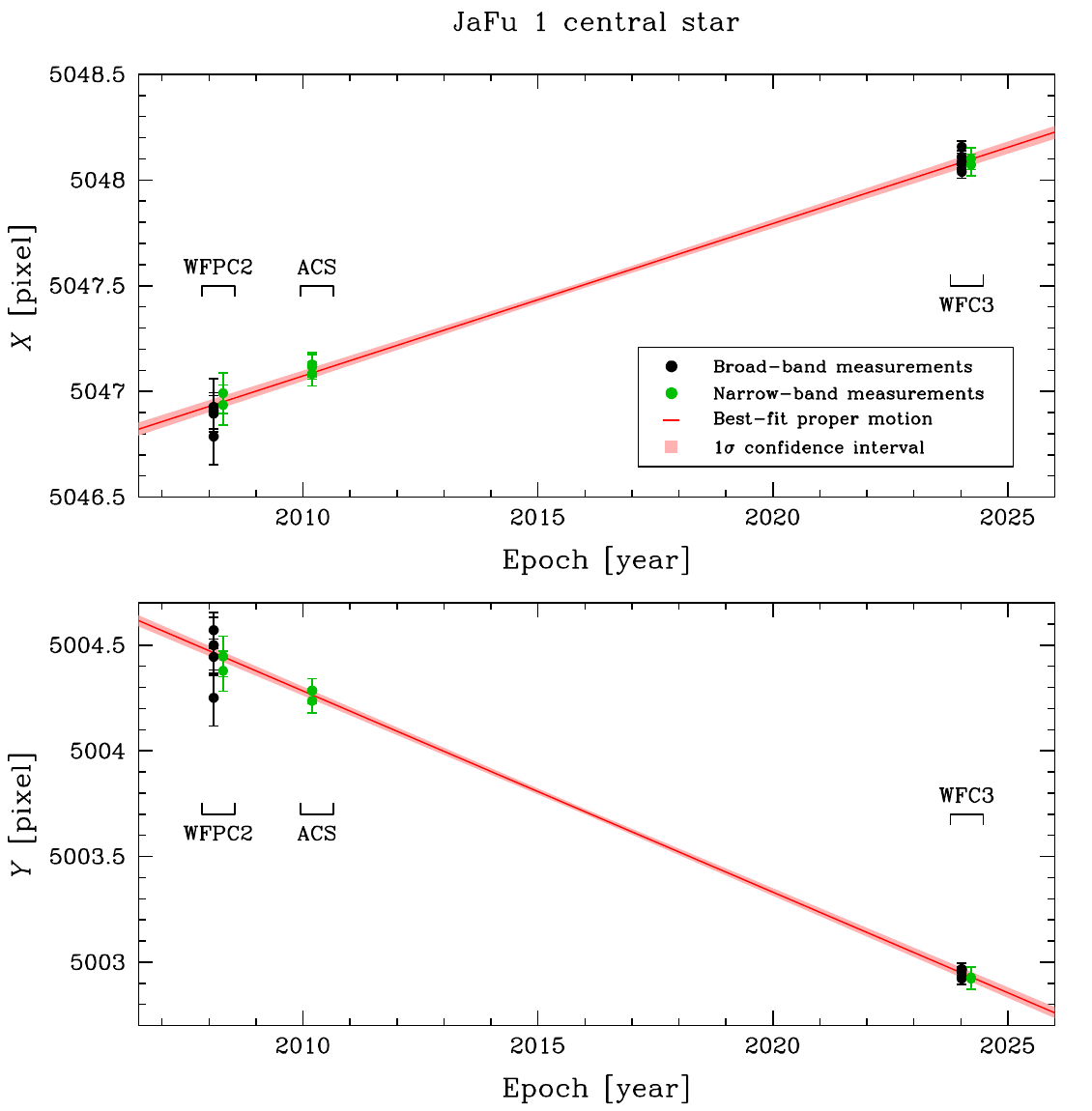}
\caption{
Measured positions of the central star of JaFu\,1 in our master reference frame versus observation epoch. Labels indicate the \HST\/ cameras used for the images. The $X$ and $Y$ positions are measured in pixels, with $X$ pointing west and $Y$ pointing north, and the pixel scale is 40~mas\,pixel$^{-1}$. Positions from broad-band filters are plotted with black filled circles, and measurements from narrow-band emission-line filters with green filled circles. Points at the same epochs are slightly separated in time for clarity. The red lines show the proper-motion fits, and the shaded red areas their 1$\sigma$ confidence intervals. See text for further details.
\label{fig:jafu1_pm} }
\end{figure*}

Figure~\ref{fig:jafu1_pm} illustrates the PM solution for the central star of JaFu\,1. Here we plot the measured positions of the star in the master reference frame at our three observation epochs; the $(X,Y)$ positions are in pixels, with $X$ toward west and $Y$ toward north. For reasons discussed in Section~\ref{sec:JaFu_as_EGB6}, we distinguish between measurements in broad-band filters (black points) and in narrow-band emission-line filters (green points), but the agreement between them is good. The red lines show linear fits to the data, whose slopes we convert to PMs using the frame scale of 40~mas\,pixel$^{-1}$. For our final PM solution, we omitted the two WFC3 2024 exposures in F656N, due to the lack of a good distortion model in that filter. However, we did determine and plot the positions in this filter by applying the distortion model for the F606W filter, and there appear to be no large discrepancies.

The PMs resulting from our analysis are relative to the average PM
of stars in the \textit{HST\/} field. To convert them into absolute
PMs, we applied a correction based on the absolute PMs of stars in common between the
\textit{HST\/} and \Gaia\/ catalogs. The mean
absolute PM components of { 122 \Gaia\/ stars  in the \HST\/ field brighter than $G=19$ }are
$(\mu_{\alpha} \,\cos\delta, \mu_{\delta})_{0} = (-2.16 \pm 0.13$, $-4.55 \pm
0.10)$ mas\,yr$^{-1}$.

Table~\ref{tab:parameters} lists our results for the nucleus of JaFu\,1. The first two rows list the central star's J2000 equatorial coordinates (epoch 2016), and the next two rows give the corresponding Galactic longitude and latitude. Rows five and six list the absolute PM components and their internal errors. In addition to these internal uncertainties there are slightly larger external errors due to the conversion from relative to absolute PMs, as presented in the previous paragraph. The final three rows list magnitudes (Vega scale) that we determined from the ACS and WFC3 frames.\footnote{We measured count rates in 10-pixel-radius apertures, corrected the rates to infinite apertures, and applied Vega-scale zero-points. The aperture corrections and zero-points for ACS and WFC3 are available from  \url{https://acszeropoints.stsci.edu} and \citet{Calamida2021}, respectively.} Calibrated magnitudes from the earlier WFPC2 frames were published by \citet{Jacoby2017}. We note that the central star is unusually bright in F502N and F656N, indicating that it has strong nebular emission lines of [\oiii] and \Ha; we return to this point in Section~\ref{sec:JaFu_as_EGB6}.

\begin{deluxetable}{lc}
\tablecaption{Parameters for the Central Star of JaFu~1  \label{tab:parameters} }
\tablehead{
\colhead{Parameter}
&\colhead{Value}
}
\decimals
\startdata
RA (J2000) & 17 43 57.239 \\
Dec (J2000) & $-26$ 11 53.83 \\
$l$ [deg] &  2.142 \\
$b$  [deg] &  +1.745 \\
$\mu_\alpha\cos\delta$ [mas\,yr$^{-1}$] & $-5.054 \pm 0.075$ \\
$\mu_\delta$ [mas\,yr$^{-1}$] & $-8.323 \pm  0.059$ \\
F502N magnitude & $21.54\pm0.01$ \\
F555W magnitude & $22.50\pm0.02$ \\
F656N magnitude & $16.77\pm0.03$ \\
\enddata
\end{deluxetable}

\section{Proper-Motion Results}

\begin{figure*}
\centering
\includegraphics[width=\linewidth]{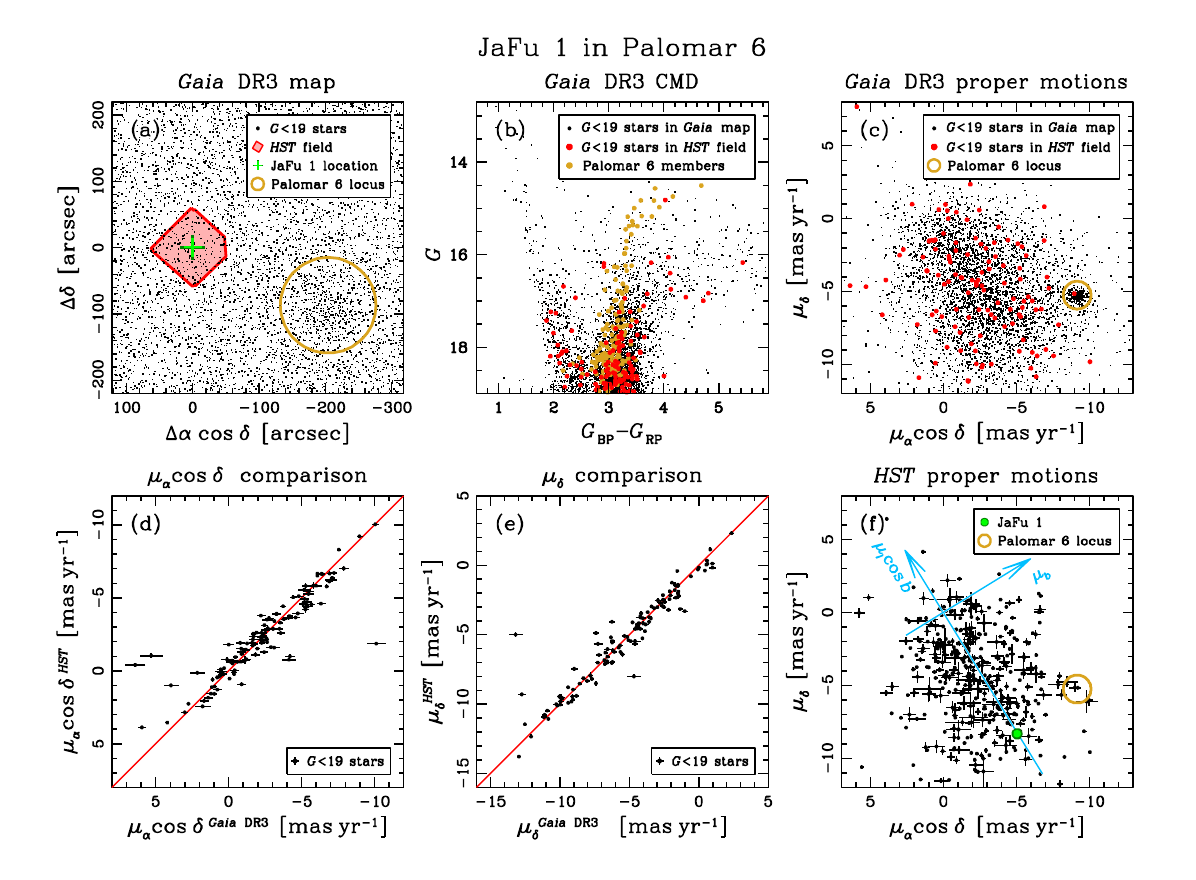}
\vskip-0.2in
\caption{
(a) Positions of \Gaia\/ DR3 stars brighter than $G=19$ in the vicinity of the globular cluster Pal\,6 (gold circle). The red border outlines the field in which archival \HST\/ data overlap with our new 2024 images. The position of the planetary nebula JaFu\,1 is marked with a green cross. 
(b)~Color-magnitude diagram for the \Gaia\/ stars. Gold points are likely cluster members, having positions and PMs within the gold circles in both Panels~(a) and~(c). Red points are the \Gaia\/ stars that lie within the \HST\/ field in Panel~(a).
(c)~Vector-point PM diagram of \Gaia\/ stars shown in Panel~(a). Red points are \Gaia\/ stars lying within the \HST\/ field of view. The gold circle highlights likely Pal\,6 members according to their PMs. 
(d)~and (e) Absolute PMs in right ascension and declination for stars measured in the \HST\/ field and cataloged in \Gaia\/ DR3. Red lines are the loci of equality. There is good consistency within the errors. (f) Vector-point diagram for the absolute PMs we measured in the \HST\/ field. The PM of the nucleus of JaFu\,1 is marked with a green circle; its uncertainties are smaller than the plotting symbol. The gold circle for the cluster motion is the same as in Panel~(c). Turquoise arrows show the directions of PMs in Galactic longitude $l$ and latitude $b$. See text for details of these diagrams.
\label{fig:jafu1_summary} }
\end{figure*}

Figure~\ref{fig:jafu1_summary} summarizes the results of our astrometric analysis. Panel~(a) presents a map showing the positions of all
\textit{Gaia\/} DR3 stars brighter than $G=19$ mag in the field of Pal~6 and JaFu~1. The gold circle at the right is centered on the cluster and has a radius of $72''$,
equal to the cluster's half-light radius \citep{Harris2010}. The red polygon northeast of the cluster outlines the field in which the
archival \HST\/ data overlap with our new 2024 data. 
A green cross at its center marks the
location of the nucleus of JaFu~1. 

Panel (b) in Figure~\ref{fig:jafu1_summary} plots the color-magnitude diagram of all the stars from
Panel (a) in the \Gaia\/ photometric system of $G$ magnitude versus $G_{\rm BP}-G_{\rm RP}$ color. \textit{Gaia} sources lying within the
half-light radius of the cluster {\it and\/} having a PM consistent with cluster membership [i.e., lying within the gold circle in Panel~(c)] are plotted with gold filled circles. Note the well-defined red-giant branch, and the lack of a blue horizontal branch, consistent with the cluster's relatively high metallicity of $\rm[Fe/H]=-0.91$ \citep{Harris2010} or slightly higher \citep[e.g.,][]{Vasquez2018}. Stars falling
within the \textit{HST\/} field are plotted using red filled circles.  

Panel~(c) shows the  vector-point 
diagram for the \Gaia\/ DR3 stars in Panels (a) and (b). The gold circle is centered on the mean PM for Pal~6 members
of
$(\mu_{\alpha} \,\cos\delta, \mu_{\delta}) = (-9.17, -5.26 )$ mas\,yr$^{-1}$, and it has a radius of 0.95 mas\,yr$^{-1}$, equal to 5 times the cluster's central
PM dispersion of 0.19 mas\,yr$^{-1}$. Here the mean PM and dispersion are both taken from \citet{Baumgardt2019}. 
Red points again correspond to stars lying in the \textit{HST\/}
field. 
It is
clear, from Panels~(b) and~(c), that most if not all of the
\textit{Gaia\/} stars located within the \textit{HST\/} field are not cluster
members: the red-giant branch of Pal~6 shown by the gold points in Panel~(b) is almost entirely missing among the red points, and there is a significant population of foreground Galactic-disk main-sequence stars. Only one \Gaia\/ star in the \HST\/ field shares the PM of Pal~6.

Comparisons between the absolute PMs of stars as measured from \textit{HST\/}
and \textit{Gaia\/} data are shown in Panels~(d) and~(e) for the right-ascension and
declination directions, respectively. PM uncertainties for the \textit{HST\/} measurements are
typically smaller than the size of the plotting symbols, but for some of the \Gaia\/ measurements the error bars are larger and are plotted. Most of the large outliers in Panels~(d) and~(e) are seen to have close neighbors at \HST\/ resolution. The red diagonal
lines are not fits to the data but simply the bisectors of the
plane, and indicate a generally good agreement between the two
sets of PM measurements (after application of the conversion from relative to absolute for the \HST\/ data, as described above).


Finally, Panel~(f) plots the vector-point diagram for the stars within the red field in Panel~(a), determined from the \HST\/ data. Here we show all measured stars with a PM error less than 2~mas\,yr$^{-1}$. The uncertanties are again usually smaller than the plotting symbols, except for a few of the fainter objects for which the error bars are shown. Of the 430 stars measured in the \HST\/ frames, only five had PM uncertainties larger than $\pm$2~mas\,yr$^{-1}$. In this panel turquoise arrows mark the directions of Galactic longitude and latitude. Note that the PM distribution is elongated in the direction of Galactic longitude, and that the Sun's participation in Galactic rotation produces a reflex offset of the mean PM from zero along this direction.  The gold circle in Panel~(f), repeated from Panel~(c), shows the locus of PMs for members of Pal~6.
Our measured absolute PM of the nucleus of JaFu~1 (listed in Table~\ref{tab:parameters}) is plotted as a green circle in Panel~(f). Its uncertainties are smaller than the plotting symbol.  



Our results show that the PM of the JaFu~1 central star differs from the mean PM of members of Pal~6 with very high statistical significance. The offset in total PM between them is 5.13~mas\,yr$^{-1}$; this is formally about a 40$\sigma$ difference, given the uncertainty in the total PM dominated by the external error of $\sim$0.13~mas\,yr$^{-1}$ (see Section~\ref{sec:dataanalysis}).


The large difference in PMs between the PN and the cluster appears to rule out cluster membership. This makes the close agreement in RVs between the two (Section~\ref{sec:JaFu1}) a remarkable coincidence---unless we consider an alternative possibility that the progenitor star was a cluster member, which has recently been ejected from the cluster through dynamical interactions or tidal stripping. However, the relative directions of the PMs make this scenario unlikely. The PN lies to the northeast of the cluster (see Figure~\ref{fig:panstarrs}). Thus, if the PN had been ejected, its PM should lie in a northeast direction relative to that of the cluster; actually, its relative PM is toward the southeast [Figure~\ref{fig:jafu1_summary}, Panel~(f)]. 


\goodbreak\smallskip 

\section{J\MakeLowercase{a}F\MakeLowercase{u}~1 as an EGB~6-type Planetary Nebula \label{sec:JaFu_as_EGB6} }

\citet{Jacoby2017} found that the central star of JaFu~1 is unusually bright in [\oiii] and \Ha, based on their photometric analyses of the 2008 and 2010 WFPC2 and ACS images.\footnote{This feature appears to confirm that the star is indeed the true nucleus of the PN, and not a chance alignment, in spite of its remarkably off-center location.} Our magnitude measurements in these bandpasses, derived from ACS and WFC3 images, confirm the star's enhanced brightness in the narrow-band filters, as shown in our Table~\ref{tab:parameters}. These results imply that JaFu~1 belongs to the rare class of ``EGB~6-type'' PNe, whose central stars are accompanied by compact emission knots (CEKs), unresolved from the nuclei in ground-based imaging. This property of the prototype, EGB~6 itself, was discovered by \cite{Ellis1984}. \citet{FrewParker2010} defined the EGB~6 class, pointing out a handful of other PNNi that are also accompanied by CEKs. \citet{Liebert2013} and \citet{BondEGB62016}, using \HST\/ images, demonstrated the remarkable fact that in EGB~6 the CEK is resolved from the central star at a separation of about $0\farcs16$, and appears to be associated with a dust-embedded low-luminosity binary companion star. More recently, \citet{Bond_Abell57_2024} have discovered that the nucleus of the PN Abell~57 is accompanied by a CEK and shares many of the properties of EGB~6, but in this case high-resolution imaging is lacking.

The similarity of the central star of JaFu~1 to the nucleus of EGB~6 raises the possibility that the former may likewise exhibit an offset in the position of the central star between narrow- and broad-band images. The distance of EGB~6, based on its \Gaia\/ trigonometric parallax, is about 750~pc. JaFu~1 is $\sim$10 times further away; if its nucleus is a binary system similar to that of EGB~6, it would not appear resolved in our \HST\/ images. However, there could potentially be an astrometric offset of as much as $\sim$16~mas of the centroid of the emission knot from that of the central star, if JaFu\,1 has a CEK separated from the central star by the same projected linear amount as in EGB\,6. This would introduce systematic errors into a PM measurement based on a combination of narrow- and broad-band images obtained at different epochs.

We carried out two tests of this possibility. First, we added measurements of the two WFC3 F656N frames from 2024 to the PM solution based on 15 frames presented above. This astrometry has relatively large systematic uncertainties, because we had to apply the distortion model for the broad-band F606W filter to these frames, given the lack of a suitable model for the narrow-band filter. The result was a change of the derived PM of the central star by less than $0.001\,\rm mas\,yr^{-1}$. Second, we measured the offset between the positions of the central star at the 2024 epoch in the two filters, F555W and F656N, again using the F606W distortion model for the \Ha\ frames. Here we found a shift of $1.0\pm0.9$~mas. Thus we do not find a statistically significant offset of the CEK from the central star in JaFu\,1, and hence no appreciable effect on our determination of its PM\null. The projected physical separation of the knot and central star in JaFu\,1 appears to be considerably smaller than in EGB\,6.



%
%
%

\section{Discussion}

We have used multi-epoch high-resolution \HST\/ imaging to measure a precise PM for the central star of the PN JaFu~1, a candidate member of the GC Pal~6. Our results reveal a large discrepancy between the PMs of the nucleus and of the cluster, decisively ruling out cluster membership. 
The close agreement in RV between the PN and cluster, discussed in Section~\ref{sec:JaFu1},  thus becomes a statistically remarkable chance coincidence.

As Panel~(f) in Figure~\ref{fig:jafu1_summary} shows, JaFu\,1 has a large negative PM in the direction of Galactic longitude. This property indicates that the object belongs to the Galactic bulge population, since Galactic-disk stars lying in the direction of the Galactic center have generally smaller PMs; see, for example, Figure~8 in \citet{Libralato2021}. The high RV of the object also supports bulge membership.

Although we conclude that JaFu\,1 is not a member of Pal~6, we are still left with three cases of PNe that appear to be genuine members of Galactic GCs, as described in Section~\ref{sec:intro}. These objects continue to challenge our understanding of the evolution of low-mass stars \citep[e.g.,][]{Jacoby2017, Bond2020, Kwitter2022}. And  JaFu~1, even if it belongs to the Galactic-bulge population rather than Pal~6, still confronts us with the puzzles of its off-center nucleus and of the nature of EGB~6 central stars.

\acknowledgments

Based on observations with the NASA/ESA {\it Hubble Space Telescope\/} obtained from the Data Archive at the Space Telescope Science Institute (STScI), which is operated by the Association of Universities for Research in Astronomy, Incorporated, under NASA contract NAS5-26555. Support for Program number GO-17531 was provided through grants from STScI under NASA contract NAS5-26555.

This work has made use of data from the European Space Agency (ESA) mission
{\it Gaia\/} (\url{https://www.cosmos.esa.int/gaia}), processed by the {\it Gaia\/}
Data Processing and Analysis Consortium (DPAC,
\url{https://www.cosmos.esa.int/web/gaia/dpac/consortium}). Funding for the DPAC
has been provided by national institutions, in particular the institutions
participating in the {\it Gaia\/} Multilateral Agreement.

The Pan-STARRS1 Surveys (PS1) and the PS1 public science archive have been made possible through contributions by the Institute for Astronomy, the University of Hawaii, the Pan-STARRS Project Office, the Max-Planck Society and its participating institutes, the Max Planck Institute for Astronomy, Heidelberg and the Max Planck Institute for Extraterrestrial Physics, Garching, The Johns Hopkins University, Durham University, the University of Edinburgh, the Queen's University Belfast, the Harvard-Smithsonian Center for Astrophysics, the Las Cumbres Observatory Global Telescope Network Incorporated, the National Central University of Taiwan, the Space Telescope Science Institute, the National Aeronautics and Space Administration under Grant No.\ NNX08AR22G issued through the Planetary Science Division of the NASA Science Mission Directorate, the National Science Foundation Grant No. AST-1238877, the University of Maryland, E\"otv\"os Lor\'and University (ELTE), the Los Alamos National Laboratory, and the Gordon and Betty Moore Foundation.




\bibliography{PNNisurvey_refs}

\end{document}